\documentclass[12pt]{iopart}
\usepackage{graphicx} 
\usepackage{cite}
\usepackage{amssymb,amsfonts, esint}
\usepackage{algpseudocode}
\usepackage{graphicx}
\usepackage{textcomp}
\usepackage{xcolor}
\usepackage{algorithm}
\usepackage{bbm}
\def\BibTeX{{\rm B\kern-.05em{\sc i\kern-.025em b}\kern-.08em
    T\kern-.1667em\lower.7ex\hbox{E}\kern-.125emX}}

\begin{document}

\title[Planar Coil Optimization for the Eos Stellarator using Sparse Regression]{Planar Coil Optimization for the Eos Stellarator using Sparse Regression}

\author{Ryan Wu}

\address{Kearny, New Jersey, USA}
\ead{ryan.wu@thea.energy}

\author{Thomas Kruger}

\address{Kearny, New Jersey, USA}

\author{Charles Swanson}

\address{Kearny, New Jersey, USA}
\vspace{10pt}
\begin{indented}
\item[]January 2025
\end{indented}

\begin{abstract}
A challenge in the design of stellarators for confining plasma at conditions relevant to fusion energy generation is designing a feasible set of magnetic field coils which can create the necessary confining field. One active direction of investigation involves the creation of a set of simplified planar coils to approximate the desired magnetic field, split between large plasma encircling coils which generate the majority of the field and small shaping coils near the plasma surface which correct remaining errors in the field geometry. The problem of optimizing currents in these coils is inherently ill-posed due to the Biot-Savart Law. In this work, the problem of optimizing the current distribution of the shaping coils is posed as a sparse regression to minimize coil count while targeting $B_n = \langle\mathbf{B} \cdot \mathbf{n}\rangle$, a metric of confinement quality, as a least-squares objective. The goal is to improve manufacturability and conductor efficiency by using fewer coils at higher currents. Using an ensemble of sparse optimization algorithms such as LASSO, relax-and-split, and heuristic methods, Pareto fronts between coil sparsity and $B_n$ properties are identified. Differences between the optimization algorithms are evaluated, demonstrating up to 20\% reductions in mean $B_n$ at the same sparsity compared to a previously used heuristic method. The perturbation sensitivity of the sparse solutions to manufacturing misalignment is also evaluated. The results from the study demonstrate the potential and diversity of sparse optimization strategies in stellarator coil design problems.
\end{abstract}

\maketitle

\section{Introduction}\label{sec:Introduction}
Recent developments in stellarator theory, optimization techniques, and experimental results demonstrated at projects such as Wendelstein 7-X have created a resurgence in the stellarator approach to magnetically confined fusion. \cite{gates_recent_2017} Stellarators do not rely upon plasma current to generate the rotational transform necessary for confining particle drifts, rather externally creating it via a twisted steady-state magnetic field. \cite{gates_stellarator_2018, helander_theory_2014} This reduces the severity of potentially damaging plasma current-driven disruptions, as the confinement field is generally stable irrespective of the plasma behavior, enabling long-duration steady operation relevant for power-generation applications. \cite{helander_stellarator_2012} However, due to the complex shape of the magnetic field, modular stellarator coils have been expensive to manufacture, assemble, and perform metrology upon, often accounting for more than half the cost of experiments and driving significant delays due to manufacturing challenges associated with tolerances on field geometry. \cite{neilson_lessons_2009, bosch_lessons_2011} 

Novel approaches to stellarator designs have sought to remedy this by shifting the complexity from manufacturing to controls and optimization, creating stellarator designs which rely upon many planar coils or permanent magnets which are relatively simple to fabricate. \cite{gates_stellarator_2024, kruger_planar_2024, qian_simpler_2022, zhu_pm4stell_2022} By decomposing the magnetic field into individual contributions from simpler coil geometries, it is possible to approximate the field created by much more complex coil structures while preserving robust maintenance and mass-manufacturing capabilities. This modification introduces additional dimensionality to the optimization problem, with the numerous individual magnets requiring tuning in placement and field strength. This is an ill-posed problem, as the Biot-Savart law dictates that exact inversion between a desired magnetic field and a set of coil currents is often not feasible while multiple combinations of coil currents may produce similar averaged field errors. \cite{landreman2017improved} One approach to this problem by Kaptanoglu et. al \cite{kaptanoglu_permanent-magnet_2022, kaptanoglu_sparse_2023} formulates the optimization as a sparse regression, minimizing the $l_0$ norm of the total number of magnets using a relax-and-split algorithm with regularization. In doing so, the authors were able to explicitly minimize the magnet count, while still optimizing for the reconstruction accuracy of the desired magnetic field. 

Convex planar coils are able to attain greater fields than permanent magnets of comparable size, and still benefit from cost reductions compared to modular coil designs due to being able to be wound in tension. Sparsity is an attractive attribute for small planar dipole coil (hereafter referred to as shaping coil) arrays as it optimizes for lower coil counts with more current in individual coils, which are easier to manufacture and cheaper in terms of system integration requirements. 

Past work in stellarator coil optimization has examined Pareto frontiers trading off between coil length in modular coilsets and magnetic field properties such as quadratic flux \cite{bindel2023understanding} or quasiaxisymmetry error \cite{giuliani2024direct}. This Pareto analysis provides a useful framework for sparse optimization as well to examine the effect of coil sparsity on field reconstruction error on the plasma surface. From these data, plant designers can identify multiple design points to balance engineering cost and vacuum vessel port accessibility with physics properties.

This work will present optimization results from extending the permanent magnet sparse regression approach to planar coil stellarators using several sparsification methods. As the shaping coils are able to continuously scale in field strength rather than the binary behavior of permanent magnets, other techniques beyond $l_0$ norm relax-and-split were used, including $l_1$-regularized LASSO regression, heuristic deletion methods based on coil current magnitudes, and mixed-integer quadratic program formulations using commercial optimization software. Results were compared in terms of Pareto optimality on the tradeoff between magnetic field reconstruction accuracy on the plasma surface and coil sparsity. 

The implementation and results from this sparse regression approach are described in the remainder of the paper. Sections \ref{sec:Method} and \ref{sec:algos} overview the plant design and shaping coil initialization and sparsification approaches. Section \ref{sec:Results} presents the results of those optimizations, including the Pareto front trading off between coil sparsity and normal magnetic field on the plasma surface for each method implemented. Sections \ref{sec:FutureWork} and \ref{sec:Conclusion} describe future directions of study and conclude the paper.

\section{Model} \label{sec:Method}
\subsection{Stellarator Model}
\begin{figure}[htbp]
\centerline{\includegraphics[width=0.8\textwidth]{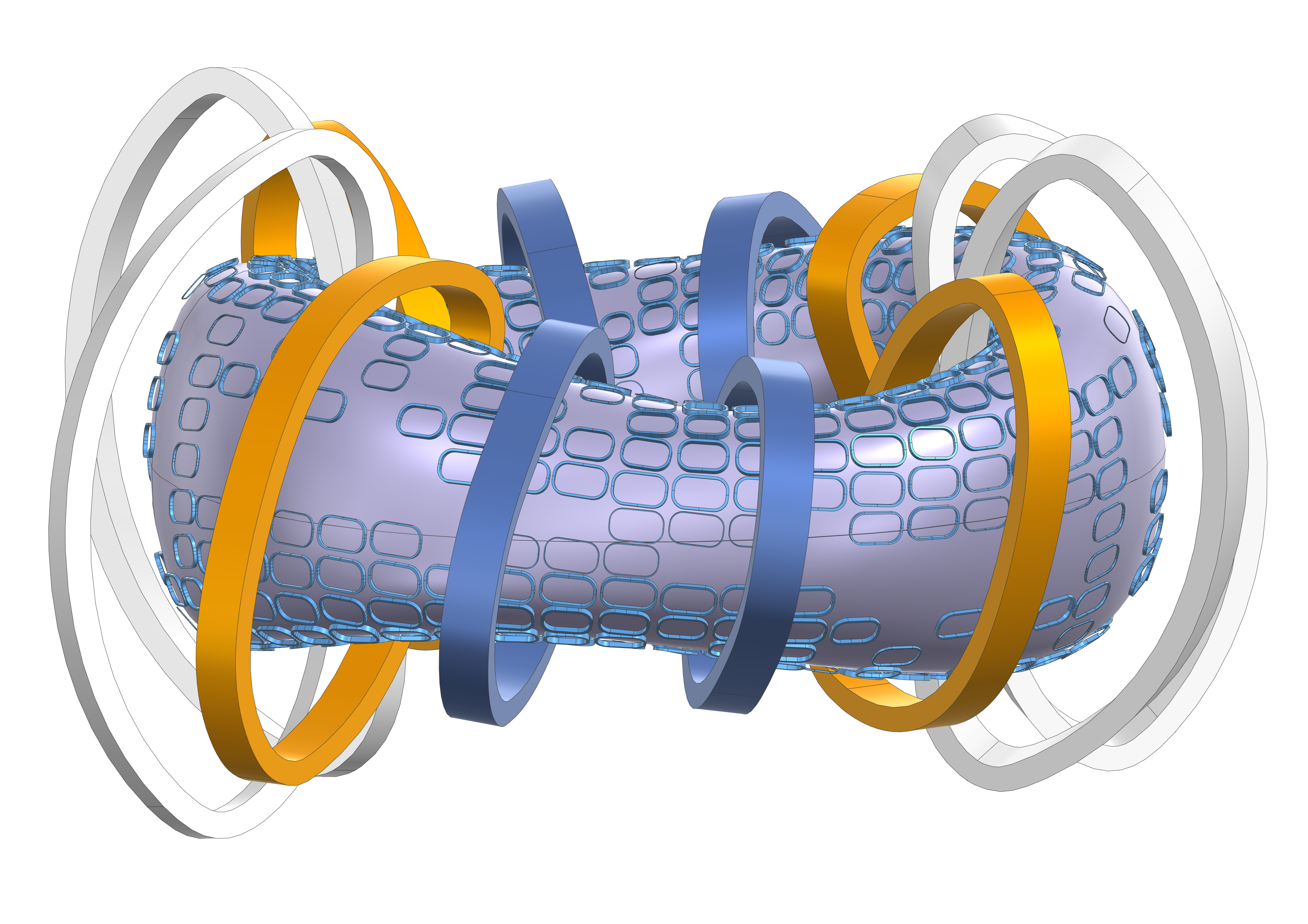}}
\caption{Rendering of Eos planar coil stellarator design. Encircling coils are in dark blue, orange, and white. Shaping coils are in light blue. The shaping coils depicted have been sparsified for heating and diagnostic access.}
\label{TheaEos}
\end{figure}

The optimization was undertaken on Thea Energy's Eos stellarator prototype design (figure \ref{TheaEos}). \cite{swanson_scoping_2024} The plasma equilibrium is quasiaxisymmetric with two field periods, spanning a major radius of 2.7 meters and an aspect ratio (torus major/minor radius) of 6:1. The baseline plant design uses an all-planar coil approach, with 12 planar toroidal coils encircling the plasma generating the majority of the confining field while an array of small shaping coils offset 50 cm off of the plasma surface provides the correction field. This combined design enables sector maintenance capabilities and relatively simple manufacturing and assembly, as the shaping coils have standardized geometry, and can be removed with segments of the stellarator wall in the gaps between the encircling coils.

\begin{figure}[htbp]
\centerline{\includegraphics[width=0.9\textwidth]{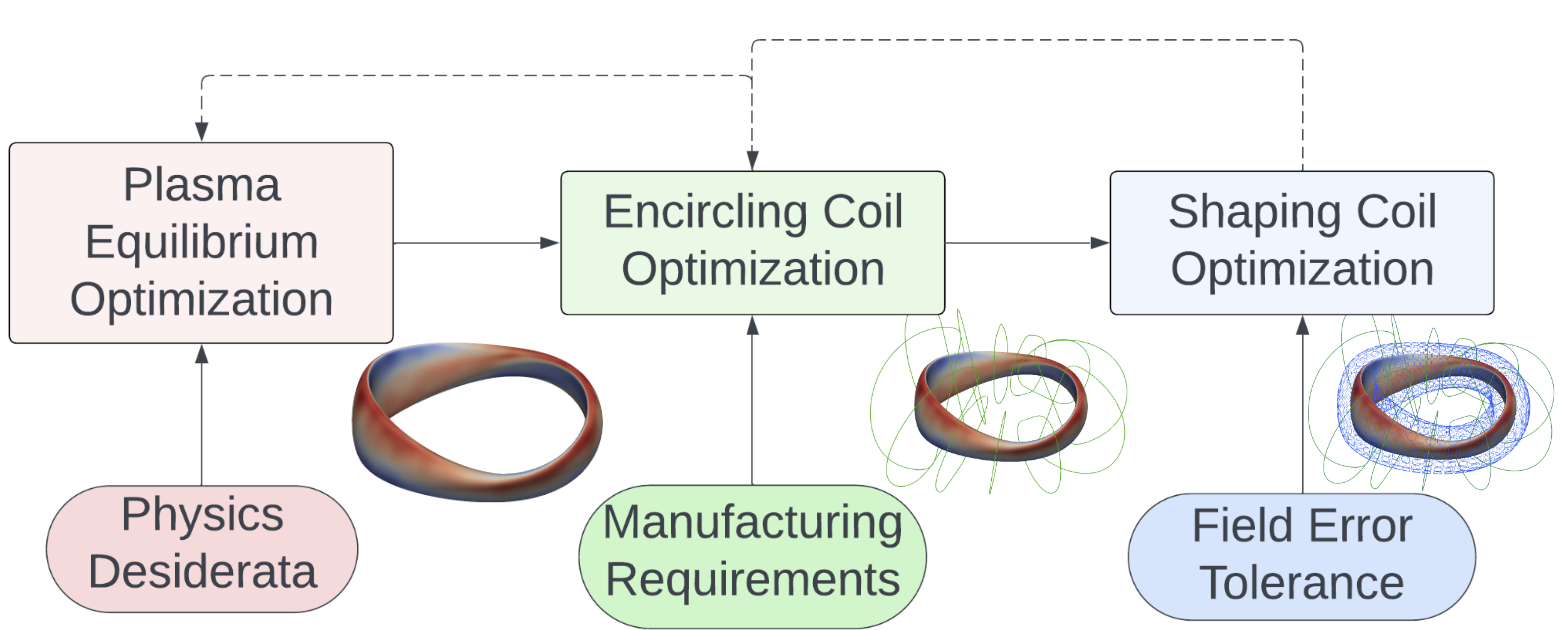}}
\caption{Three-stage optimization for the planar coil stellarator. The forward (left to right, solid) arrows represent the initial optimization steps to generate the shaping coil set from the plasma equilibrium, and the backward (dashed) arrows represent additional free-boundary and coil optimizations which are initialized with a shaping coil solution.}
\label{TheaOptimization}
\end{figure}

\begin{figure}[htbp]
\centerline{\includegraphics[width=0.6\textwidth]{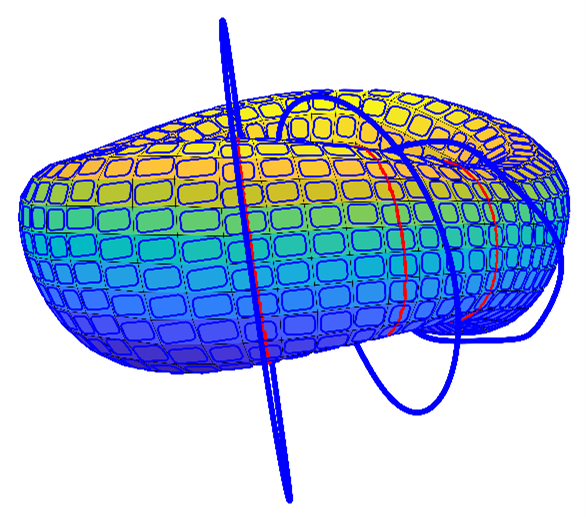}}
\caption{Encircling coil geometry overlaid on the shaping coil array. The encircling coils define cut-planes (red lines) on a winding surface. The cut-planes are used to orient poloidal columns of shaping coils.}
\label{ToroidalOnly}
\end{figure}

The optimization of the design proceeds in three steps (figure \ref{TheaOptimization}). The plasma equilibrium is first optimized for confinement quality, followed by optimization of the encircling coils to minimize normal field $B_n = \langle\mathbf{B} \cdot \mathbf{n}\rangle$ on the plasma surface. These first two optimization steps can be iteratively refined to create equilibria which are amenable to simpler coil geometries. Finally, the shaping coils are placed and their currents are optimized with a least-squares $B_n$ as the cost metric. This final optimization step is the primary target of this project, with existing plasma equilibria geometry and encircling coils (figure \ref{ToroidalOnly}) imported from previous optimization results \cite{kruger_planar_2024} used to create the field which must be error-corrected. In this study, the only free parameters are the currents in the shaping coils, while the geometry of all coils, plasma equilibrium, and encircling coil currents are held constant.

Though this is not as strongly ill-posed as the permanent magnet model with tens of thousands of candidate positions in layers of magnets, the proximity and number (O($10^3$)) of candidate shaping coils still allow numerous valid current distributions to approximate the same field profile. Once a shaping coil solution which satisfies engineering requirements is identified, that solution is used to seed additional optimizations with encircling coil geometry also free to change as well as free-boundary equilibria optimization. The objective of the work presented in this paper will be to find sparse shaping coil solutions for initializing downstream all-coil and free-boundary optimizations.

Initial optimization of the encircling coils without the shaping coils present is performed in FOCUS with the space curve parameterization constrained to a plane. \cite{zhu_new_2017, kruger_planar_2024} The DESC stellarator optimization package is used for importing and computing the equilibrium plasma surface geometry, as well as the encircling coil contribution to the field on the plasma surface. \cite{panici2023desc} 

\subsection{Shaping Coil Initialization}
A winding surface is used to orient and locate the shaping coils into an array, using the method described in \cite{kruger_planar_2024}. The winding surface used in this work is a 50 cm uniform expansion of the plasma boundary along the surface normals. Encircling coil planes are used to define initial cut-planes on the winding surface. These cut-planes are interpolated to create additional columns of rounded rectangular shaping coils (as seen on figure \ref{ToroidalOnly}). Shaping coil centroids are evenly distributed around the columns. There are 12 poloidal columns in each half field period, 3 of which are aligned with the encircling coils. This design uses 20 shaping coils per poloidal column, each having the same poloidal dimension, for a total of 240 shaping coils per half field period. The toroidal dimensions of the shaping coils are chosen from a list of 5 different sizes to maximally cover the winding surface with coils while still maintaining a small number of unique shaping coil sizes. The shaping coil orientation is determined by the normal direction of the winding surface and the tangent vector to the encircling coil cut planes.

\begin{figure}[htbp]
\centerline{\includegraphics[width=0.6\textwidth]{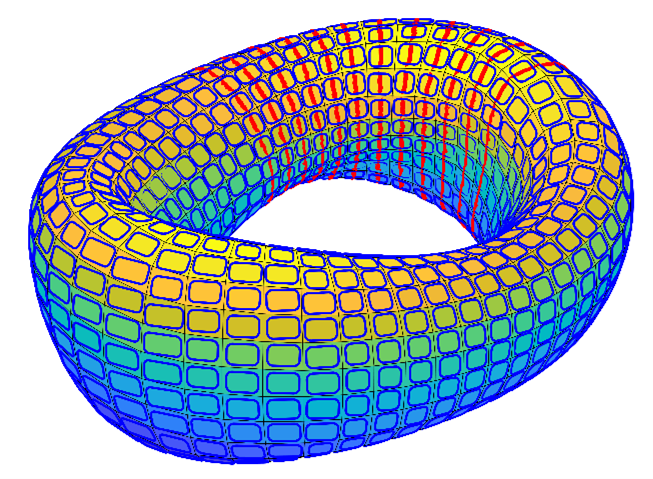}}
\caption{Candidate shaping coil locations on plasma offset surface. Cut-planes (red lines) are interpolated to define additional poloidal columns of shaping coils. This grid uses one column of shaping coils directly beneath each encircling coil on the cut-planes, and three columns of shaping coils between the cut-planes.}
\label{DipoleLocations}
\end{figure}

\subsection{Shaping Coil Optimization Formulation}
The problem of minimizing $B_n$ over the plasma surface (PS) is realized in the form of $\min \oiint_{PS} |\mathbf{B\cdot n}|^2 ds$, which can be discretized over a fixed set of quadrature points on the PS as $\min \sum_\mathbf{x} |\mathbf{B(x)\cdot n(x)}|^2$. The magnetic field component $\mathbf{B_{shaping_1}(x)\cdot n(x)} \propto i_1$ for a single coil current $i_1$ due to linearity in the Biot-Savart Law for fixed coil geometry and quadrature point geometry. This linearity is captured in an induction matrix $\mathbf{A}$ relating the current vector $\mathbf{i}$ of the shaping coils and the normal component of the  magnetic field at the quadrature points. To account for stellarator symmetry, each column of $\mathbf{A}$ encodes the normal magnetic field at all quadrature points from a given coil and all symmetric copies of that coil, allowing the optimization and sparsification to preserve stellarator and field period symmetries. In summary, the matrix element $\mathbf{A_{jk}}$ is the proportionality constant between $|\mathbf{B(x_j)\cdot n(x_j)}|$ and the shaping coil current $i_k$.

The generated shaping coilset was imported into DESC, and a Fourier planar coil representation (determined by center, normal vector, and Fourier series in radius) was fit to each coil in the coilset. A linear grid of 65 poloidal $\times$ 65 toroidal quadrature points was sampled from a half field period portion of the plasma boundary, accounting for stellarator symmetry, and the corresponding surface area of each point was computed as $\mathbf{s} = |\mathbf{e}_\theta \times \mathbf{e}_\zeta|$ to weight each quadrature point. The quadrature point grid was chosen to be significantly more dense than the shaping coil grid to accurately capture field ripple induced by the shaping coils. Virtual casing was performed on the equilibrium to find the magnetic field from plasma currents $\mathbf{B_{plasma}}$ on those quadrature points. The magnetic fields from the encircling coil-only optimized solution on the quadrature points $\mathbf{B_{encircling}}$ was also computed.

The optimization objective at the quadrature points becomes $\mathbf{b} = -(\mathbf{B_{plasma}} + \mathbf{B_{encircling}}) \cdot \mathbf{n} $, where $\mathbf{b}$ is a vector of the same dimensionality as the number of quadrature points and $\mathbf{n}$ are the plasma surface normal vectors at each of those quadrature points. The least-squares optimization problem is now 

\begin{equation}
\label{eq:baseopt}
    \min_{\mathbf{i}} f_{flux}(\mathbf{i}) = ||(\mathbf{Ai}-\mathbf{b}) \circ \mathbf{s}||_2 \\
    \textrm{s.t.} ||\mathbf{i}||_\infty \leq i_{max}.
\end{equation}

The weighting is applied to each quadrature point with the Hadamard product $\circ$, and the current in each coil is limited by engineering limits to $i_{max} = $ 1.3 MA. $\mathbf{A}$ is a matrix of dimension $4225$ (number of quadrature points) $\times$ 240 (number of candidate shaping coils in one half field period). Though the problem described in equation \ref{eq:baseopt} is convex, previous work in permanent magnet optimization \cite{kaptanoglu_permanent-magnet_2022} has noted that the objective landscape near the global minimum can be described as locally flat, admitting many solutions of similar $f_{flux}$ but very different current distributions.

\section{Optimization Algorithms}\label{sec:algos}
\subsection{Optimize then Delete (OtD) Algorithm}
Previous work on sparsifying the shaping coil set used a heuristic approach which iteratively removed the lowest-current shaping coils. \cite{kruger_planar_2024} This is similar to a sequentially thresholded least squares scheme which has been used elsewhere in sparse regression. \cite{brunton2016discovering} After solving the least-squares regression problem in equation \ref{eq:baseopt}, an additional constraint was added that set the lowest-current coil to exactly zero current, and the process was repeated.

\begin{algorithm}
\caption{Optimize then Delete}
\begin{algorithmic}[1]
    \State Initialize $\mathcal{C} = \{\}$ 
    \For{$n = 1$ to $N$} 
        \State Solve equation \ref{eq:baseopt} with additional constraints $\mathcal{C}$ to find $\mathbf{i}^*$
        \State $j \gets \arg\min_k |i^*_k|$ 
        \State Add constraint $i_j = 0$ to $\mathcal{C}$
    \EndFor
    \State \Return $\mathbf{i}^*$ 
\end{algorithmic}
\end{algorithm}

\subsection{Randomized OtD}
Additionally, it is valuable for downstream engineering planning to carry forward multiple shaping coil patterns at the same sparsity.  OtD is unable to do this, since the solution is deterministic at each sparsity level. A modified version of OtD was implemented to rectify this, which removes coils with a probability proportional to the magnitude of their current. To generate the dataset, 100 runs of randomized OtD was performed, which was comparable in runtime to the mixed-integer quadratic programming approach below.

\begin{algorithm}
\caption{Randomized OtD}
\begin{algorithmic}[1]
    \State Initialize $\mathcal{C} = \{\}$ 
    \For{$n = 1$ to $N$} 
        \State Solve  equation \ref{eq:baseopt} with additional constraints $\mathcal{C}$ to find $\mathbf{i}^*$
        \State Compute probabilities $p_k = \frac{\max(\mathbf{|i|}^*) - |i^*_k|}{\sum_{j} \big(\max(\mathbf{|i|}^*) - |i^*_j|\big)}$ for all $k$
        \State $j \gets $randomly select index based on probabilities $p_k$
        \State Add constraint $i_j = 0$ to $\mathcal{C}$
    \EndFor
    \State \Return $\mathbf{i}^*$ 
\end{algorithmic}
\end{algorithm}

\subsection{LASSO}
A commonly used method in sparse regression is LASSO regression, which adds a regularized $l_1$ norm to the objective for variable selection. \cite{tibshirani1996regression} Due to the shape of the $l_1$ norm ball, LASSO regression can explicitly set terms of the $\mathbf{i}$ vector to zero. The Langrangian form of the regression problem is used:

\begin{equation}
\label{eq:lassoopt}
    \min_{\mathbf{i}} f_{flux}(\mathbf{i})  + \beta||\mathbf{i}||_1 \\
    \textrm{s.t.} ||\mathbf{i}||_\infty \leq i_{max}.
\end{equation}

$\beta$ controls the strength of the regularization, and 100 values of $\beta$ from $10^{-5}$ to $10$ were used, which generated solutions for the full range of 0 to 100 \% sparsity.

\subsection{Relax-and-Split}
Explicitly optimizing for sparse solutions requires adding an $l_0$ regularization term, which introduces nonconvexity into the problem. Zheng and Aravkin propose a relax-and-split algorithm \cite{zheng_relax-and-split_2020} which is provably convergent for such objective functions. The problem is formulated as:

\begin{equation}
\label{eq:rsopt}
    \min_{\mathbf{i}} f_{flux}(\mathbf{i})  + \alpha||\mathbf{i}||_0 \\
    \textrm{s.t.} ||\mathbf{i}||_\infty \leq i_{max},
\end{equation}

where $\alpha$ is a parameter which controls the strength of the $l_0$ norm regularization. The relax-and-split algorithm uses an auxilliary variable $\mathbf{w}$ which is least-squares constrained with $\mathbf{i}$ to create an alternating minimization scheme. $\mathbf{w}$ contains solely the nonconvex component of the optimization without the $f_{flux}(\mathbf{i})$ contribution, which can be efficiently computed using the proximal gradient. The $\mathbf{i}$ is then solved for fixed $\mathbf{w}$, and the process repeats until convergence. \cite{kaptanoglu_permanent-magnet_2022}

\begin{algorithm}
\caption{Relax-and-Split}
    \begin{algorithmic}
        \Require $0 \leq \alpha_j \leq 1, \nu$
        \For{$\alpha_j \in {\alpha_{series}}$}
            \For{$n = 1$ to $N$ }
                \State $\mathbf{i}_n \gets argmin_{\mathbf{i}}[\frac{f_{flux}(\mathbf{i})^2}{2} + \frac{||\mathbf{i}-\mathbf{w}_{n-1}||_2^2}{2\nu}]$
                \State $w_i \gets prox_{\alpha_j, l_0}(\mathbf{i}_n)$
            \EndFor
            \State $w_0 \gets w_{maxIter}$
        \EndFor
    \end{algorithmic}
\end{algorithm}

The initial value chosen for $\mathbf{w}_0$ was the zero vector. It was noted in previous research and replicated in this study that the $\alpha$ $l_0$ normalization works best when gradually tightened in an iterative manner, where each proximal solution was used to initialize the next iteration. This was implemented in the study, and improved convergence stability. Hyperparameter grid sweeps were performed over $\alpha$ (0.01-0.6), $\nu$ ($10^{-1} - 10^{7}$), and $\mathrm{len}(\alpha_{series}) \in \{1,2,5\}$. The PermanentMagnet subroutines in SIMSOPT are used for the relax-and-split optimization. \cite{bharat_medasani_hiddensymmetriessimsopt_2023, kaptanoglu_permanent-magnet_2022}

\subsection{Mixed-Integer Quadratic Programming (MIQP)}
It is also possible to formulate this sparse regression as an MIQP, where an integer sparsity vector $\mathbf{z}$ explicitly constrains which indices of the current vector $\mathbf{i}$ are nonzero.

\begin{equation}
\label{eq:miqpopt}
    \min_{\mathbf{i}, \mathbf{z}} f_{flux}(\mathbf{i}) \\
    \textrm{s.t.} \quad 
    \mathbf{z}^\top \mathbf{1} = p - N, \\
    \mathbf{|i|} \leq i_{max} \mathbf{z},  \quad
    \mathbf{z} \in \{0, 1\}^p.
\end{equation}

$N$ sets the number of removed shaping coils and $p$ is the total number of candidate shaping coils. It is intractable to prove global optimality in this problem for large $N$ and $p$, as the number of possible sparsity patterns $\mathbf{z}$ grows combinatorially like $p \choose N$. However, commercial MIQP solvers exist that provide heuristic solutions with low computational cost. For this study, the Gurobi optimizer \cite{gurobi} was used, which implements several proprietary heuristic methods to find high-quality solutions to the problem in equation \ref{eq:miqpopt}, even if strictly proving optimality is computationally prohibitive. The optimization was restricted to 15 minutes on 8 threads of a laptop i7-1370P processor for each sparsity level $N$.

\subsection{Analysis Procedure}
To compare all results based solely on the quality of the identified shaping coil sparsity pattern $\mathbf{z}$, the solution $\mathbf{i}$ from each optimization algorithm was masked such that $\mathbf{z} = \mathbbm{1}(\mathbf{i} \neq 0)$, where $\mathbbm{1}$ is the indicator function. $\mathbf{z}$ is then used to initialize a masked version of the optimization problem: 

\begin{equation}
\label{eq:maskedopt}
    \min_{\mathbf{i}^*} f_{flux}(\mathbf{i}^*) \\
    \textrm{s.t.} \quad 
    \mathbf{|i^*|} \leq i_{max} \mathbf{z}.
\end{equation}

This avoids issues where the regularization method affects the current on coils which are not removed, which is particularly relevant in the LASSO penalty case. All optimization results were then compared using $B_{n,avg} = \frac{1}{||\mathbf{s}||_1}||(\mathbf{Ai}^*-\mathbf{b}) \circ \mathbf{s}||_1$, as well as $B_{n,max} = \max(\mathbf{Ai}^*-\mathbf{b})$. Additionally, differences in sparsity patterns $n_{diff} = ||\mathbf{z_1}-\mathbf{z_2}||_1$ were examined, which is useful for informing the degree of flexibility available to engineers when downselecting coil configurations. For solutions where the optimization approach generates several candidate solutions at the same sparsity level, such as from randomized OtD or the hyperparameter sweep for relax-and-split, the Pareto front between sparsity and the desired parameter is examined.

\section{Results} \label{sec:Results}
\subsection{Pareto Fronts}
\begin{figure}[htbp]
\centerline{\includegraphics[width=0.95\textwidth]{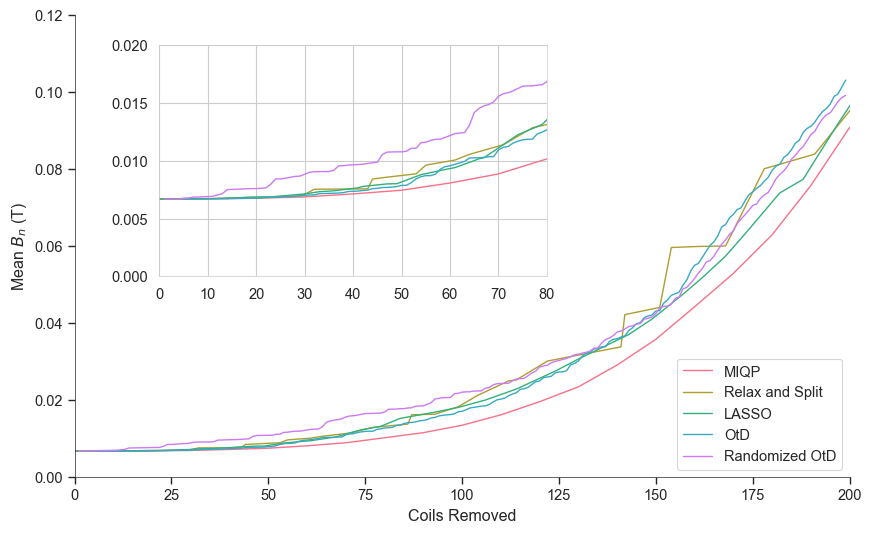}}
\caption{Pareto front of average $B_n$ over the plasma surface at different sparsity levels for each optimization method. The inset contains a close-up of the same data within the range of 0 to 80 coils removed.}
\label{ParetoAverage}
\end{figure}

\begin{figure}[htbp]
\centerline{\includegraphics[width=0.6\textwidth]{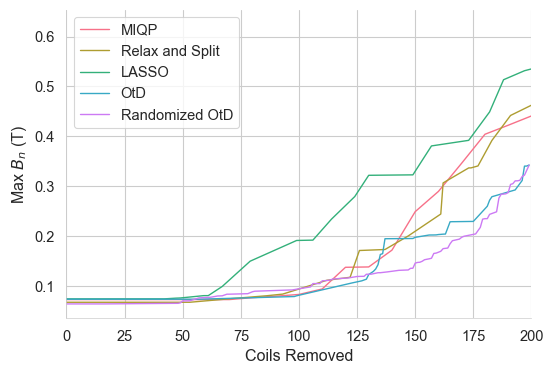}}
\caption{Pareto front of maximum $B_n$ over the plasma surface at different sparsity levels for each optimization method.}
\label{ParetoMax}
\end{figure}

The Pareto tradeoff front between coil sparsity and normal magnetic field on the plasma surface was examined for all optimization algorithms used. Figures \ref{ParetoAverage} and \ref{ParetoMax} show the results for mean and maximum $B_n$ respectively, demonstrating different performance characteristics for each optimization method at different sparsity levels. The MIQP approach has lower mean $B_n$ than all other optimization approaches at every sparsity level. Of the remaining approaches, OtD performs well at low sparsity (up to 125 out of 240 coils removed), but at higher sparsity randomized OtD and LASSO exhibit lower mean $B_n$. Results in max $B_n$ are more mixed, and no definitive winner emerges in the low sparsity regime, though this metric is more sensitive to grid parameterization over the plasma surface. However, LASSO in particular appears to perform poorly in max $B_n$.

\begin{figure}[htbp]
\centerline{\includegraphics[width=0.6\textwidth]{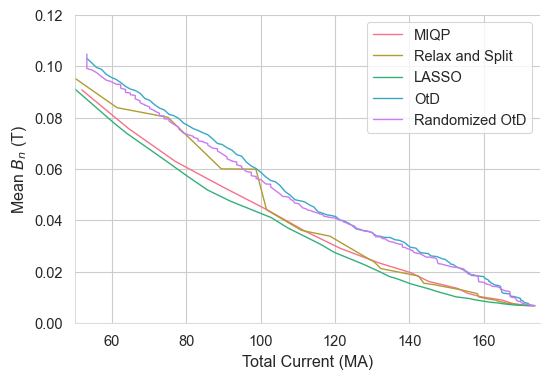}}
\caption{Pareto front of mean $B_n$ over the plasma surface at different total shaping coil currents for each optimization method. The total current in the shaping coils is for one half-field period only, not accounting for the stellarator and field period symmetric copies of the coils.}
\label{ParetoCurrent}
\end{figure}

A useful proxy for the superconductor usage of the stellarator shaping coils is the total current, which has an approximate correspondence when neglecting the effects of background field. The Pareto front between total current and mean $B_n$ is depicted in figure \ref{ParetoCurrent}. At every value of total current, LASSO finds the coil solution with the minimum mean $B_n$ of the optimization methods tested. This is in line with expectations, as the $l_1$ norm penalty in LASSO effectively includes total current within the optimization objective, though it is not necessarily a given since the current has been reoptimized with the masked problem in equation \ref{eq:maskedopt}. OtD yielded the greatest mean $B_n$ at all current values, which may be attributable to how it iteratively selects for high-current coils with the deleting operation.

\begin{figure}[htbp]
\centerline{\includegraphics[width=0.6\textwidth]{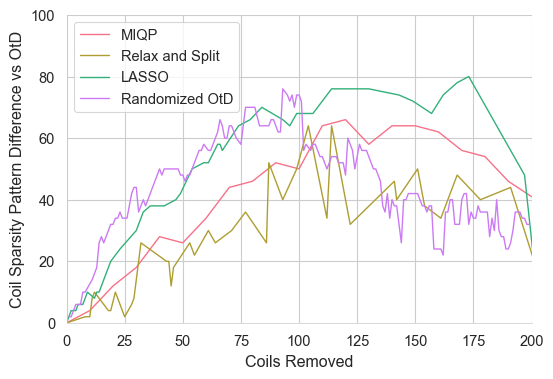}}
\caption{Difference in sparsity patterns $n_{diff} = ||\mathbf{z_1}-\mathbf{z_{OtD}}||_1$ plotted between Pareto-optimal mean $B_n$ solutions (same as figure \ref{ParetoAverage}) and the OtD solution at the corresponding sparsity level.}
\label{SparsityDiffs}
\end{figure}

To verify that multiple sparsity patterns are actually present in the different Pareto fronts, the total difference $n_{diff} = ||\mathbf{z_1}-\mathbf{z_{OtD}}||_1$ between each Pareto-optimal sparsity  vs. mean $B_n$ solution and the number of shaping coils removed is plotted in figure \ref{SparsityDiffs}. In the 0 to 100 coils removed regime, the increasing $n_{diff}$ with sparsity indicates that significantly different solutions are being found by the other optimization algorithms. However, these solutions are not completely different from OtD, as the maximum possible $n_{diff}=240$ at 120 shaping coils removed (inverse of the OtD solution) is not reached, indicating that there are certain indices of $\mathbf{z}$ preserved in multiple Pareto-optimal solutions. This result motivates further analysis to identify these preserved coils and their role within the optimization and engineering design process.

\subsection{Sample Solutions}
For additional insight, Pareto-optimal solutions with 80 coils removed using MIQP, LASSO, and OtD are examined (Table \ref{tabsparse80}). The MIQP-optimized solution reduces mean $B_n$ by 20\% compared to the OtD solution at the same sparsity, and has approximately the same mean $B_n$ as the Pareto-optimal OtD solution with only 64 shaping coils removed in the half field period. 

Meanwhile, the 80 coil removed LASSO solution has a total current which is 12.6\% lower than the OtD solution with a similar mean $B_n$ (at 84 coils removed), and has a mean $B_n$ 51\% lower than the OtD solution with the same total current (at 125 coils removed).

\begin{table}[htbp]
\begin{center}
\begin{tabular}{l|lll}
      & Mean $B_n$ (T)  & Max $B_n$ (T)  & Total Current (MA) \\ \hline
OtD   & 0.0127          & 0.088          & 165.7              \\
LASSO & 0.0134          & 0.148          & \textbf{144.7}     \\
MIQP  & \textbf{0.0102} & \textbf{0.076} & 159.2             
\end{tabular}
\caption{Characteristic quantities for optimization solutions with 80 coils removed in one half field period.}
\label{tabsparse80}
\end{center}
\end{table}

\begin{figure}[htbp]
\centerline{\includegraphics[width=0.85\textwidth]{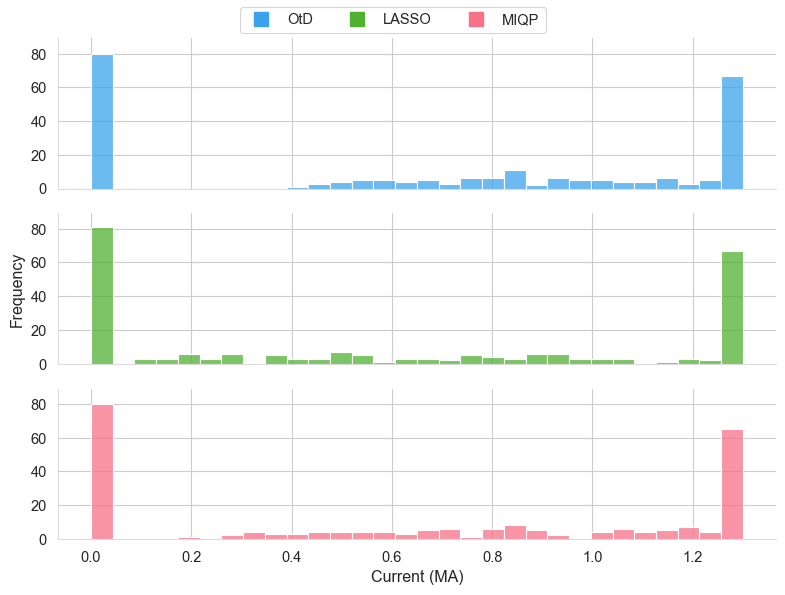}}
\caption{Histogram of current in shaping coils in the Pareto-optimal 80 coils removed solution for the OtD, LASSO, and MIQP optimization methods.}
\label{Histogram}
\end{figure}

\begin{figure}[htbp]
\centerline{\includegraphics[width=0.99\textwidth]{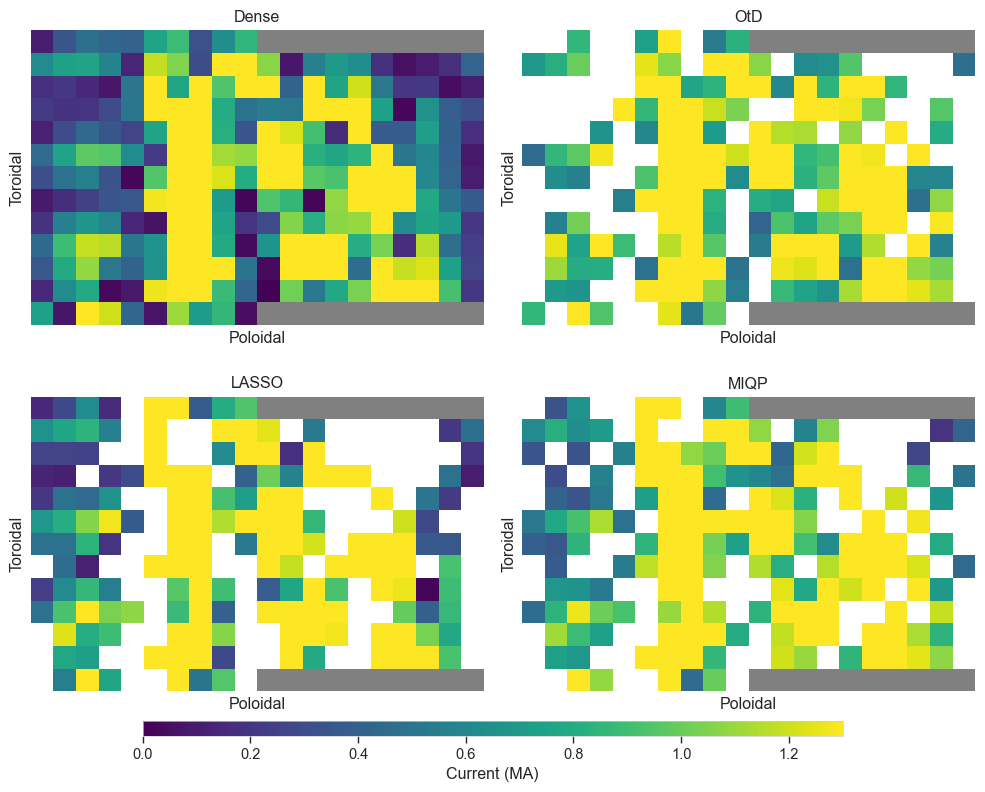}}
\caption{Grids representing currents in each candidate shaping coil oriented in toroidal and poloidal directions within one half field period: dense coil grid with no removals (top left), OtD 80 removed (top right), LASSO 80 removed (bottom left), MIQP 80 removed (bottom right). Coils with zero current are represented in white. Null entries due to coils at the edges of the half field period lying on the symmetry plane are represented in gray. The top (positive z coordinate) edge of the plasma cross section roughly corresponds to the 1/3 column in the poloidal axis, and exhibits consistently high shaping coil currents.}
\label{DipoleMatrix}
\end{figure}

The current distribution in the shaping coils for the three optimization methods at this sparsity is plotted in figure \ref{Histogram} and visualized in figure \ref{DipoleMatrix}. As previously discussed, the OtD method appears to be generating solutions with shaping coils at higher currents compared to the other two optimization approaches, consistent with the nature of the low-current deletion heuristic. In the process, large swaths of coils are removed (eg. near the top left of the OtD heatmap in figure \ref{DipoleMatrix}). In contrast, LASSO produces coil distributions with much lower currents, again consistent with the $l_1$ penalty. The MIQP solution, though it had better mean $B_n$ than the other optimizers, also had a broad distribution of coil currents with one coil having only 0.2 MA, indicating that these low-current coils are important for reducing $B_n$, a phenomenon which would not have been identified using the thresholding method of OtD. This is also visible in the checkerboard coil pattern visible throughout the MIQP heatmap in figure \ref{DipoleMatrix}, which may help to smooth out the normal field distribution over the plasma surface compared to the large deleted blocks from OtD. 

\begin{figure}[htbp]
\centerline{\includegraphics[width=0.95\textwidth]{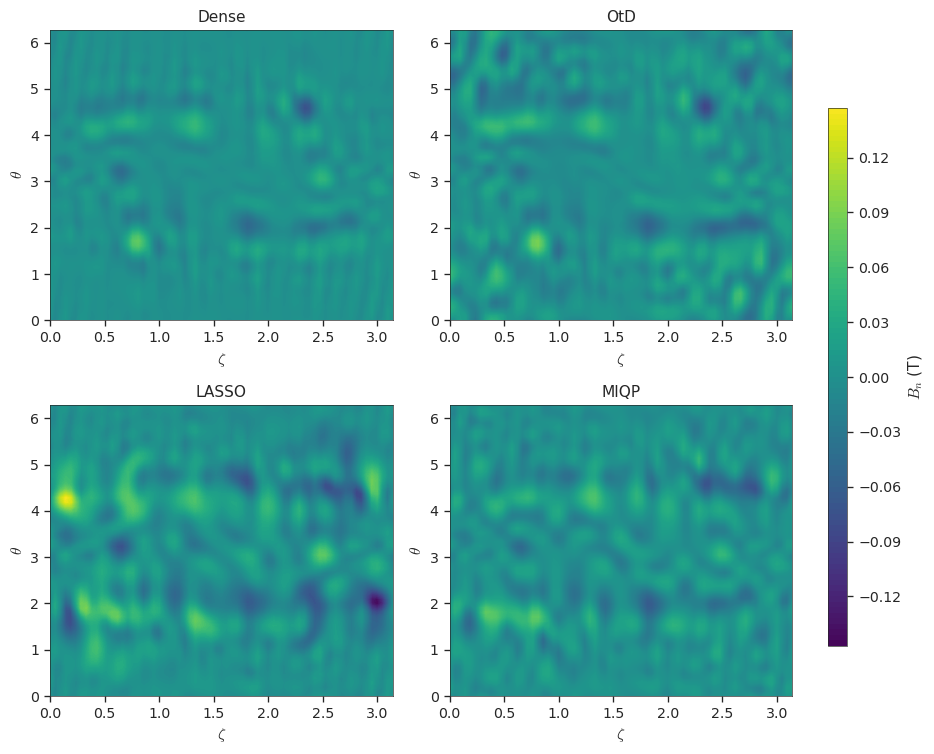}}
\caption{Contour maps of $B_n$ on the plasma surface from the dense 0 removal grid (top left) and three optimized solutions with 80 coils removed. The contours are plotted in toroidal-poloidal ($\zeta, \theta$) coordinates.}
\label{BnContours}
\end{figure}

\begin{figure}[htbp]
\centerline{\includegraphics[width=0.9\textwidth]{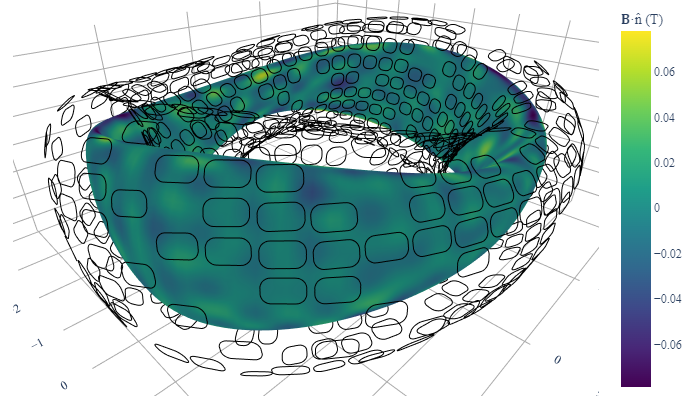}}
\caption{Contour map of $B_n$ on the plasma surface and sparsified coilset for the MIQP 80 coils removed solution.}
\label{Coils3D}
\end{figure}

Examining the $B_n$ distributions on the plasma surface also shows differences between the optimization solutions (figures \ref{BnContours} and \ref{Coils3D}). The LASSO solution has a significantly larger range of $B_n$ variability over the plasma surface compared to the OtD, MIQP, and unsparsified (dense) coilsets. The OtD and MIQP solutions add smaller-scale structures to the $B_n$ contour compared to the dense solution, in the form of ripples where coils have been removed. Additional work is required to further understand the effects of these $B_n$ changes on confinement properties of the free-boundary equilibria.

\subsection{Perturbation Robustness}
The sensitivity of the shaping coil optimization to perturbations in placement was also examined. Due to tolerances within the manufacturing and assembly process, errors will exist in the placement of the shaping coils, manifesting analytically as a $\mathbf{\Delta A}$ deviation in the $((\mathbf{A+\Delta A})\mathbf{i}-\mathbf{b})\circ \mathbf{s}$ objective term, since the position change modifies the induction coefficient between the coil currents and the plasma surface quadrature points. To quantify this effect, a 1 cm perturbation in a random direction was applied to the centroid of each shaping coil. Additionally, the normal vector of the coil was rotated by 1 degree about a random orthogonal vector to the normal. Both perturbations are on the order of worst-case expected manufacturing and mounting tolerances.

\begin{figure}[htbp]
\centerline{\includegraphics[width=0.7\textwidth]{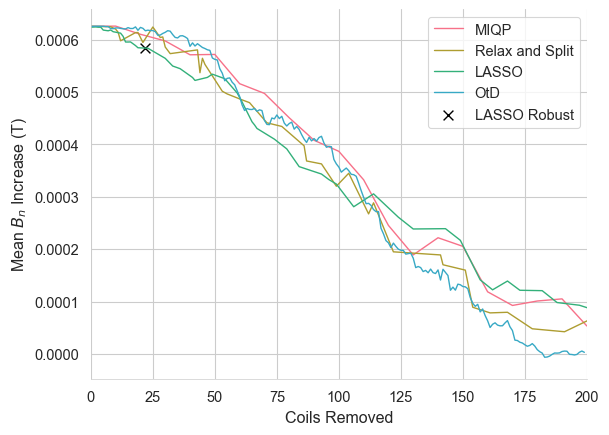}}
\caption{Perturbation sensitivity of sparsity-optimized configurations with random shaping coil displacements and rotations. The configurations shown are the same Pareto-optimal sets from figure \ref{ParetoAverage}, and the black cross is a LASSO solution derived from the Xu et. al \cite{xu_robust_2010} robust optimization formulation. Mean $B_n$ increases are measured relative to the unperturbed $B_n$ results for the same shaping coil configuration.}
\label{Perturbation}
\end{figure}

The average mean $B_n$ change in the perturbed cases compared to the unperturbed case (averaged over 50 sets of perturbations) is displayed in figure \ref{Perturbation}. Currents were held constant in the coils, assuming that the modification to the induction matrix was not identified in operation and reoptimization was not performed. All optimization methods shown demonstrate a decrease in mean $B_n$ change with increasing sparsity, which is consistent with the fact that there are fewer coils to perturb at those higher sparsity levels. LASSO appears to perform slightly better than the other methods within the regime with less than 100 coils removed.

In past work, Xu et. al \cite{xu_robust_2010} have demonstrated that in the case of robust least-squares regression with feature-wise independent disturbances, the problem reduces to a LASSO $l_1$ norm constrained optimization. Specifically, given a problem of the form $\min_\mathbf{i}\{ \max_{\mathbf{\Delta A}} \{ ||\mathbf{b}-(\mathbf{A+\Delta A})\mathbf{i}||_2 \}\}$ with a maximum feature-wise disturbance $\mathbf{\Delta A} \in {\mathbf{\delta_j}}$ bounded by $||\mathbf{\delta_j}||_2 \leq c_j$, an equivalent regression problem is $\min_\mathbf{i} \{ ||\mathbf{b}-\mathbf{A}\mathbf{i}||_2 + \sum_j c_j |i_j| \}$, which is simply a case of a LASSO optimization. The perturbations presented here are indeed feature-wise independent, as each perturbation is applied on the individual coil level and only affects the row of the induction matrix $\mathbf{A}$ associated with that coil. $c_i$ is also bounded given a bound on the perturbation magnitude, though this bound is difficult to calculate analytically due to the relationship of coil geometry, perturbation, and quadrature point locations.

An attempt at estimating this $\mathbf{c}$ vector was undertaken from measuring the maximum feature-wise perturbation magnitudes on the coilset $\mathbf{A + \Delta A}$ matrices, which was used to initialize an additional robust LASSO run with $\mathbf{c}$ replacing the weight scalar. The result is plotted on figure \ref{Perturbation}, and is on the Pareto front for the LASSO method as well, indicating that both approaches are finding similar solutions. The solution is not the absolute best in terms of mean or max $B_n$ over the perturbed dataset, but is within 2\% and 10\% of the best values found from the optimization solution set, for each parameter respectively. More investigation is still necessary to explicitly bound perturbation magnitudes $c_i$ for the optimized results, but the result demonstrates how the sparse optimization techniques employed in this work could have implications for robustness as well. 

\section{Future Work} \label{sec:FutureWork}
Further investigation of the model will involve examining the assumptions regarding shaping coil location and finite build geometry. The sparse shaping coil grid can be used in the initialization of downstream optimization in a code such as FOCUS or DESC, where coil position and angle can be further optimized considering additional objectives such as forces, torques, and critical current. Additionally, restrictions on coil placement from maintainability, diagnostic placement, and actuator access also must be considered, adding constraints on invalid coil locations. 

The initial coil array over which sparsification occurs does not necessarily need to be feasible. Ongoing work aims to examine whether this approach can be extended to initial shaping coil grids with overlapping coils, where additional constraints are added to prevent intersecting coils from being present in a sparse solution. This essentially allows coil position and orientation to be wrapped into the same optimization script, at the cost of additional combinatorial complexity.

Lastly, an optimized sparse shaping coil result can inform upstream optimization steps for the equilibrium and encircling coils, particularly as the sparse coil placement indicates specific areas requiring field correction. In a combined optimization scheme, it may be possible to configure the upstream optimization solutions to produce a $B_n$ field particularly amenable to sparse coil correction, or use the robustness of the shaping coil locations to inform tolerances or constraints on encircling coil design. Work is planned to investigate modifying the quadrature node weights within the encircling coil optimization to prioritize locations where the shaping coils from the sparse solution are unable to fully correct the $B_n$.

\section{Conclusion} \label{sec:Conclusion}
Stellarators provide an attractive pathway to the realization of magnetically confined fusion for power generation, and the planar coil stellarator design holds promise in easing manufacturing requirements for the concept. This work demonstrates the  application of sparse optimization techniques to the shaping coilset for one candidate planar coil stellarator design, yielding a family of sparse shaping coil solutions with different current distributions, total currents, and $B_n$ profiles. A mixed-integer quadratic program formulation is shown to identify the coil sparsity pattern with the Pareto-optimal mean $B_n$ at every sparsity level tested, and a LASSO formulation identifies all Pareto-optimal sparsity patterns when trading between total current and mean $B_n$. The perturbation robustness of the optimized results was also assessed, and found to be consistent with theory on robust regression. Further work on this model will involve integration with the established Thea optimization toolchain to feed back into filamentary coil optimization processes.

\section*{Acknowledgment}

I wish to thank John Wright, Elizabeth Paul, and Alan Kaptanoglu for providing feedback and discussions on the optimization structure, hyperparameter search strategy, and further steps.

\section{References}
\enlargethispage{1\baselineskip}
\bibliography{SparsePaper}
\end{document}